\documentclass[twocolumn, nofootinbib, aps, prd, 10pt]{revtex4-1}
\usepackage{graphicx, amsmath, amssymb, bm}
\usepackage[breaklinks, colorlinks, citecolor=blue, linkcolor=blue, urlcolor=blue]{hyperref}

\newcommand{\rhct}{{R_\text{h} = ct}}

\begin{document}

\title{Robust model comparison disfavors power law cosmology}

\author{Daniel L. Shafer}
\email{dlshafer@umich.edu}
\affiliation{Department of Physics, University of Michigan, 450 Church Street, Ann Arbor, MI 48109-1040, USA}

\begin{abstract}

Late-time power law expansion has been proposed as an alternative to the standard cosmological model and shown to be consistent with some low-redshift data. We test power law expansion against the standard flat $\Lambda$CDM cosmology using goodness-of-fit and model comparison criteria. We consider Type Ia supernova (SN Ia) data from two current compilations (JLA and Union2.1) along with a current set of baryon acoustic oscillation (BAO) measurements that includes the high-redshift Lyman-$\alpha$ forest measurements from BOSS quasars. We find that neither power law expansion nor $\Lambda$CDM is strongly preferred over the other when the SN Ia and BAO data are analyzed separately but that power law expansion is strongly disfavored by the combination. We treat the ${R_\text{h} = ct}$ cosmology (a constant rate of expansion) separately and find that it is conclusively disfavored by all combinations of data that include SN Ia observations and a poor overall fit when systematic errors in the SN Ia measurements are ignored, despite a recent claim to the contrary. We discuss this claim and some concerns regarding hidden model dependence in the SN Ia data.

\end{abstract}

\maketitle

\section{Introduction} \label{sec:intro}

Despite the general observational success of $\Lambda$CDM in describing the detailed properties of the Universe and its expansion, some alternative models for expansion have been proposed. A notable alternative is power law cosmology, where the scale factor evolves purely as a function of the proper time (age) to some constant power. While constraints from Big Bang nucleosynthesis suggest that a power law model cannot describe the complete expansion history of the Universe \cite{Kaplinghat:1998wc, Kaplinghat:2000zt}, it may be more plausible as a description of a low-redshift modified-gravity alternative to the cosmological constant. For instance, power law expansion could result from a coupling of classical fields to spacetime curvature, regardless of matter content (e.g.\ \cite{Dolgov:1996zg}). Theoretical motivation aside, it is instructive to compare alternative models with $\Lambda$CDM to test the robustness of the data and their ability to discriminate between competing models.

Power law expansion has been shown (e.g.\ \cite{Jain:2003vq, Sethi:2005au, Zhu:2007tm}) to be consistent with a variety of cosmological probes for a power law exponent in the range ${1 \lesssim n \lesssim 1.5}$. More recently, \citet{Dolgov:2014faa} studied power law cosmology using current data from observations of Type Ia supernovae (SNe Ia) and baryon acoustic oscillations (BAO). They found that low-redshift power law cosmology is a good fit to the SN Ia data if the power law exponent has a value ${n \simeq 1.5}$. Here we will study similar data, but rather than simply constrain the model parameters and confirm that one can find a good fit, we would like to explicitly compare power law cosmology with $\Lambda$CDM. We will use several alternative model comparison statistics to determine if (and if so, how strongly) various data combinations prefer power law expansion over $\Lambda$CDM.

A similar alternative to $\Lambda$CDM is the so-called $\rhct$ universe proposed by \citet{Melia:2011fj}, where $R_\text{h} = c/H(t)$ is the Hubble radius. Though presented as a distinct model, the expansion history of $\rhct$ matches that of a power law with exponent ${n = 1}$ (a constant rate of expansion). Note that a constant rate of expansion has been studied earlier \cite{Dev:2000du, Dev:2002sz} under the name ``linear coasting cosmology.''

Whether or not the $\rhct$ model is unphysical has been debated in the literature. Aside from the nucleosynthesis arguments \cite{Kaplinghat:1998wc, Kaplinghat:2000zt} that apply to power law expansion in general, \citet{Lewis:2013pha} points out that the $\rhct$ universe, which is constrained to have an effective equation of state $\bar{w} = -1/3$, requires either that $\Omega_m = 0$ or that the dark energy equation of state evolves in an unphysical way at early times. In a response, \citet{Melia:2014aja} argues that the assumption of a conserved matter field is not justified even at late times, though one might view this as rather contrived. The idea of $R_\text{h}$ as a meaningful cosmic horizon has also been challenged \cite[and references therein]{Lewis:2013ks}. Setting these important concerns aside, we may still ask whether the observations favor $\rhct$ expansion as a phenomenological description of the late Universe.

An analysis by \citet{Bilicki:2012ub} determined that the $\rhct$ universe is ruled out by both SN Ia data and $H(z)$ data (from BAO and cosmic chronometers) separately, though, as pointed out in \cite{Melia:2013hsa}, the conclusions are based heavily on visual inspection of plots of reconstructed dynamical quantities. While there is nothing wrong with this dynamical approach, and while the inconsistency with $\rhct$ indeed appears quite significant, direct model comparison can quantify the preference of the data and appropriately account for differences in model complexity. That analysis also ignores systematic errors in the SN Ia measurements, and these are significant for current data. In addition, proponents of $\rhct$ cosmology have argued \cite{Melia:2012zy, Wei:2015xca} that using SN Ia data may be unfair anyway because of hidden model dependence.

In this paper, we test both power law cosmology and, separately, the $\rhct$ cosmology against $\Lambda$CDM using current data and robust model comparison statistics. The outline of the rest of the paper is as follows. In Sec.~\ref{sec:models}, we briefly review these three models for cosmic expansion. In Sec.~\ref{sec:data}, we describe the datasets we will use, and in Sec.~\ref{sec:method}, we review the goodness-of-fit and model comparison statistics. Our results are presented in Sec.~\ref{sec:results}. Finally, in Sec.~\ref{sec:discuss}, we discuss the issue of hidden model dependence in the SN Ia data, explaining why there is no real problem, before summarizing our conclusions.

\section{Models} \label{sec:models}

In this section, we briefly describe the three models we wish to compare: $\Lambda$CDM, power law cosmology, and the $\rhct$ cosmology.

\subsection{\texorpdfstring{$\Lambda$CDM}{LCDM}}

The flat $\Lambda$CDM model is considered the standard model of modern cosmology, and it is motivated by a combination of physics and empirical observations. In the $\Lambda$CDM framework, the present-day Universe consists mostly of cold dark matter and the simplest form of dark energy, the cosmological constant $\Lambda$. The $\Lambda$CDM model is usually described as having six free parameters (e.g.\ $\Omega_m$, $\Omega_c h^2$, $\Omega_b h^2$, $n_s$, $A_s$, $\tau$), remarkably simple for a model that describes the Universe as a whole (for comparison, the Standard Model of particle physics has 19 free parameters). Moreover, if the focus is only on late-time relative expansion, there is really only one free parameter, which we take to be $\Omega_m$. For a flat universe, the present cosmological-constant density is fixed to the value $\Omega_\Lambda = 1 - \Omega_m$, and the comoving angular diameter distance $r(z)$ coincides with the line-of-sight comoving distance:
\begin{align}
r(z) &= \frac{c}{H_0} \int_0^z \frac{dz'}{E(z')} \ , \\[1ex]
E(z) &\equiv \frac{H(z)}{H_0} = \sqrt{\Omega_m (1 + z)^3 + (1 - \Omega_m)} \ .
\end{align}
Note that, since we are focusing on low-redshift expansion only, we can ignore contributions to $H(z)$ from radiation or relativistic neutrinos, which have a negligible impact on expansion to the precision we are concerned with here.

\subsection{\texorpdfstring{Power law and $\rhct$ cosmology}{Power law and Rh = ct cosmology}}

In power law cosmology, the scale factor evolves with proper time (age) as
\begin{equation}
a(t) = \left(\frac{t}{t_0} \right)^n ,
\end{equation}
where $a_0 = a(t_0) = 1$ is the present value. In this case, we have
\begin{equation}
E(z) = (1 + z)^{1/n} ,
\end{equation}
so that
\begin{equation}
r(z) = \frac{c}{H_0} \times \left\{
\begin{array}{cr} \displaystyle
\frac{(1 + z)^{1 - 1/n} - 1}{1 - 1/n}, & n \neq 1 \ , \\[3ex]
\ln(1 + z), & n = 1 \ .
\end{array}
\right.
\label{eq:rzpower}
\end{equation}
Here there is one free parameter, the power law exponent $n$, which can be restricted to be in the range ${0 < n < \infty}$ if we agree we live in an expanding universe. Since the data combinations we will analyze here all exclude ${n \leq 0}$ anyway, this restriction does not affect our analysis.

The $\rhct$ universe \cite{Melia:2011fj}, though proposed as a distinct model, has an expansion history that matches that of a power law with ${n = 1}$. In the definition, $R_\text{h}$ is the gravitational radius, which is equivalent to the Hubble radius $R_\text{h} = c/H(t)$ for a flat universe, so one can also write ${Ht = 1}$ to describe this model. The comoving angular diameter distance is then given by the second case in Eq.~\eqref{eq:rzpower}. There are no free parameters in this model, so only nuisance parameters associated with the data can be varied.

Note that we are assuming a flat universe for these models as well, an assumption which can be relaxed (e.g.\ \cite{Dolgov:2014faa}). Flatness is empirically motivated (e.g.\ \cite{Ade:2013zuv}) for $\Lambda$CDM-like models, where $\Omega_k$ is constrained to be small, but the tightest constraints are somewhat model-dependent since they rely on CMB observations (see Sec.~\ref{sec:BAO}). On the other hand, if the inflationary picture is correct, we still have strong theoretical motivation to assume a flat universe, and we proceed with this assumption.

\section{Data Sets} \label{sec:data}

We now describe in some detail the data used in the analysis. It is crucial that we only choose data whose interpretation is independent of the cosmological model. To this end, we consider SN Ia observations as well as measurements of the BAO feature in large-scale structure. These are among the most mature, well-studied, and robust probes of dark energy and cosmic expansion at present. In this section and also in Sec.~\ref{sec:discuss}, we discuss some important issues related to the model-independence of these data.

We intentionally do not use some other measurements of distance and expansion rate, such as cosmic chronometers (e.g.\ \cite{Moresco:2012by, Moresco:2012jh}), which measure $H(z)$ directly by estimating the ages of passively evolving galaxies. While very promising, this method is newer and less well-studied, and the systematic errors on individual measurements are often as large as the statistical errors. In addition, we leave out measurements of the CMB distance because the measurement is at least somewhat model-dependent (see Sec.~\ref{sec:BAO}).

\subsection{SN Ia data} \label{sec:SN}

Type Ia supernovae (SNe Ia) are very bright standard candles (or \emph{standardizable} candles) that are useful for measuring cosmological distances. SNe Ia alone provided the first convincing evidence for accelerated cosmic expansion \cite{Riess:1998cb, Perlmutter:1998np}. Today, not only have we observed many more SNe, but we have also improved our understanding of their light curves and performed rigorous analyses of systematic errors (e.g.\ \cite{Conley:2011ku, Kessler:2012gn, Scolnic:2013efb, Mosher:2014gyd}). Although measurements of CMB anisotropies and large-scale structure can constrain the matter content of the Universe and even the dark energy equation of state, SNe Ia have an important role in breaking degeneracies to achieve precision constraints on dark energy.

The distance modulus of a SN at redshift $z$ is given by
\begin{equation}
\mu(z) = 5 \log_{10} \left[\frac{H_0}{c} D_L(z) \right] ,
\label{eq:mu}
\end{equation}
where ${D_L(z) = (1 + z) \: r(z)}$ is the luminosity distance. Here we have defined the distance modulus without the $H_0$ term, which is degenerate with the SN Ia absolute magnitude (see below).

Useful correlations between the peak luminosity of SNe Ia and both the stretch (or broadness) and photometric color of their light curves improve the standardization of SNe Ia by reducing the intrinsic scatter in their luminosities (and simultaneously mitigating potential systematic effects). Simply put, a broader or bluer SN light curve corresponds to a brighter SN. More recently, it has become apparent that properties of the host galaxy correlate with the intrinsic luminosity as well, and understanding these effects is the focus of much current work (e.g.\ \cite{Sullivan:2010mg, Rigault:2013gux, Rigault:2014kaa}). It is now common practice to fit for two absolute magnitudes, splitting the sample using a stellar mass cut of the host galaxy. We therefore compare the predicted distance modulus with its measured value after light-curve correction:
\begin{equation}
\mu_\text{obs} = m - (\mathcal{M} - \alpha \, s + \beta \, C + P \, \Delta M) ,
\label{eq:muobs}
\end{equation}
where $m$ is the apparent magnitude in some photometric band, $s$ and $C$ are the stretch and color measures, which are specific to the light-curve fitter (e.g.\ SALT2 \cite{Guy:2007dv}) employed, and ${P \equiv P(M_* > 10^{10} M_\odot)}$ is the probability that the SN occurred in a high-stellar-mass host galaxy. The stretch, color, and host-mass coefficients ($\alpha$, $\beta$, and $\Delta M$, respectively) are nuisance parameters that should ideally be constrained along with any cosmological parameters. The constant ${\mathcal{M} = M - 5 \log_{10} [H_0/c \times 10~\text{pc}]}$ absorbs the $H_0$ term from Eq.~\eqref{eq:mu} and is yet another nuisance parameter.

Recent analyses (see below) have concentrated on estimating correlations between measurements of individual SNe in order to appropriately account for the numerous systematic effects which must be controlled in order to improve constraints significantly beyond their current level. A complete covariance matrix for SNe Ia includes estimates of all identified systematic errors in addition to the intrinsic scatter and other statistical errors. The $\chi^2$ statistic is then calculated in the usual way for correlated measurements:
\begin{equation}
\chi^2 = \Delta \bm{\mu}^\intercal \mathbf{C}^{-1} \Delta \bm{\mu} \ ,
\label{eq:chi2}
\end{equation}
where $\Delta \bm{\mu} = \bm{\mu}_\text{obs} - \bm{\mu}(\bm{\theta})$ is the vector of residuals between the observed, corrected distance moduli and the theoretical predictions that depend on the set of cosmological model parameters $\bm{\theta}$ and $\mathbf{C}$ is the $N_\text{SN} \times N_\text{SN}$ covariance matrix for the observed distance moduli.

In this work, we use current SN Ia datasets from two alternative analyses: the joint light-curve analysis (JLA) of SNe from the Supernova Legacy Survey (SNLS) and the Sloan Digital Sky Survey (SDSS) and the Supernova Cosmology Project's Union2.1 compilation.

\subsubsection{JLA} \label{sec:JLA}

The joint light-curve analysis (JLA) \cite{Betoule:2014frx} includes recalibrated SNe from the first three years of SNLS \cite{Conley:2011ku, Betoule:2012an} as well as the complete SN sample from SDSS \cite{Sako:2014qmj}, and it is the largest combined SN analysis to date. The final compilation includes 740 SNe, $\sim$100 low-redshift SNe from various subsamples, $\sim$350 from SDSS at low to intermediate redshifts, $\sim$250 from SNLS at intermediate to high redshifts, and $\sim$10 high-redshift SNe from the Hubble Space Telescope.

We use the SN Ia data and individual covariance matrix terms provided (\url{http://supernovae.in2p3.fr/sdss_snls_jla/}) to compute the full covariance matrix, which includes statistical errors and all identified systematic errors. The covariance matrix, like the corrected distance moduli themselves, is a function of the light-curve nuisance parameters $\alpha$ and $\beta$. In the analysis, we vary all of the SN Ia nuisance parameters ($\alpha$, $\beta$, $\mathcal{M}$, $\Delta M$), recomputing the covariance matrix whenever $\alpha$ or $\beta$ is changed.

\subsubsection{Union2.1} \label{sec:Union2.1}

The Union2.1 analysis \cite{Suzuki:2011hu} from the Supernova Cosmology Project (\url{http://supernova.lbl.gov/Union/}) adds $\sim$15 high-redshift SNe to the Union2 compilation \cite{Amanullah:2010vv}, making Union2.1 the compilation with the most high-redshift SNe ($\sim$30 at ${z > 1}$) to date.

The SN distance moduli provided have been pre-corrected for stretch, color, and host-mass correlations using best-fit values for $\alpha$, $\beta$, and $\Delta M$. While we do include all identified systematic errors via the covariance matrix provided, we keep $\alpha$ and $\beta$ fixed at their best-fit values in the analysis because the covariance matrix is a function of these parameters and individual covariance matrix terms are not provided. While fixing $\alpha$ and $\beta$ is unlikely to affect the results of our model comparison (see Sec.~\ref{sec:results}), it is important to allow the effective value of the SN Ia absolute magnitude to change, so we let both $\mathcal{M}$ and $\Delta M$ vary in the analysis.

\subsection{BAO data} \label{sec:BAO}

Baryon acoustic oscillations (BAO) are the regular, periodic fluctuations of visible matter density in large-scale structure resulting from sound waves propagating in the early Universe. In recent years, precise measurements of the BAO scale at a variety of redshifts have proven to be effective probes of cosmic expansion and dark energy \cite{Weinberg:2012es, Aubourg:2014yra}. The principal observable is the \textit{ratio} of the BAO distance scale at low redshift to the comoving sound horizon ${r_d = r_s(z_d)}$ at the redshift of baryon drag (${z_d \simeq 1060}$, shortly after recombination at ${z_* \simeq 1090}$).

Typically the BAO feature is assumed to be isotropic and is identified from a spherically-averaged power spectrum. In this case, the observable is $D_V(z_\text{eff})/r_d$, where $D_V$ is a spherically-averaged (two transverse and one radial) distance measure \cite{Eisenstein:2005su} given by
\begin{equation}
D_V(z) \equiv \left[r^2(z) \, \frac{cz}{H(z)} \right]^{1/3},
\label{eq:DV}
\end{equation}
where ${r(z) = (1 + z) \: D_A(z)}$ is the comoving angular diameter distance and $H(z)$ is the Hubble parameter. More recently, it has become possible to robustly measure radial and transverse clustering separately, allowing for anisotropic BAO. In that case, the observables $r(z_\text{eff})/r_d$ and $c/(H(z_\text{eff}) \, r_d)$ are measured separately (but with some statistical correlation).

We follow \cite{Aubourg:2014yra} and combine recent measurements of the BAO feature from the Six-degree-Field Galaxy Survey (6dFGS) \cite{Beutler:2011hx}, the SDSS-II DR7 main galaxy sample (MGS) \cite{Ross:2014qpa}, and the SDSS-III Baryon Oscillation Spectroscopic Survey (BOSS) DR11 LOWZ \cite{Tojeiro:2014eea} and CMASS \cite{Anderson:2013zyy} samples. We also include a combined measurement from BOSS Lyman-$\alpha$ forest (Ly$\alpha$F) auto-correlation \cite{Delubac:2014aqe} and cross-correlation \cite{Font-Ribera:2014wya}. We use pairs of anisotropic measurements for the CMASS and Ly$\alpha$F samples and isotropic measurements for the others.

The BAO measurements used in this analysis are summarized in Table~\ref{tab:bao}. As discussed in \cite{Aubourg:2014yra}, statistical correlations (covariance) between these different samples should be negligible, so we treat them as independent in the analysis. Note that the CMASS anisotropic measurements are correlated with coefficient $-0.52$, while the Ly$\alpha$F measurements are correlated with coefficient $-0.48$.

\begin{table}
\setlength{\tabcolsep}{0.6em}
\begin{tabular}{llcr}
\hline \hline
Sample 	         & $z_\text{eff}$ & Observable  					       & Measurement        \\ \hline
6dFGS 	         & 0.106 			    & $D_V(z_\text{eff})/r_d$      & $3.047  \pm 0.137$ \\
SDSS MGS    	   & 0.15 			    & $D_V(z_\text{eff})/r_d$      & $4.480  \pm 0.168$ \\
BOSS LOWZ        & 0.32 				  & $D_V(z_\text{eff})/r_d$      & $8.467  \pm 0.167$ \\
BOSS CMASS   	   & 0.57           & $r(z_\text{eff})/r_d$        & $14.945 \pm 0.210$ \\
BOSS CMASS 	     & 0.57           & $c/(H(z_\text{eff}) \, r_d)$ & $20.75  \pm 0.730$ \\
BOSS Ly$\alpha$F & 2.34           & $r(z_\text{eff})/r_d$        & $36.489 \pm 1.152$ \\
BOSS Ly$\alpha$F & 2.34           & $c/(H(z_\text{eff}) \, r_d)$ & $9.145  \pm 0.204$ \\
\hline \hline
\end{tabular}
\caption{Summary of BAO measurements combined in this analysis. We list the sample from which the measurement comes, the effective redshift of the sample, the observable quantity constrained, and its measured value. The anisotropic measurements from BOSS CMASS are correlated with coefficient $-0.52$, while those from BOSS Ly$\alpha$F are correlated with coefficient $-0.48$. Otherwise, we assume the measurements to be statistically independent.}
\label{tab:bao}
\end{table}

The likelihood for the BAO observables is not Gaussian far from the peak. For a finite detection significance of the BAO feature, the actual likelihood will eventually asymptote to a flat tail, since any value for the observable is equally probable in the event of a non-detection \cite{Bassett:2010ku}. This is particularly important to consider when constraining parameters to a high confidence level or claiming that a model is a very poor fit to the data.

We account for this effect by applying the fitting function proposed in \cite{Bassett:2010ku} to approximate the correct likelihood. For a given signal-to-noise ratio ($S/N$), the usual ${\Delta \chi^2_\text{G} = -2 \ln \mathcal{L}_\text{G}}$ for an observable with a Gaussian likelihood is replaced by
\begin{equation}
\Delta \chi^2 = \frac{\Delta \chi^2_\text{G}}{\displaystyle \sqrt{1 + \Delta \chi^4_\text{G} \left(\frac{S}{N} \right)^{-4}}} \ .
\label{eq:chi2fit}
\end{equation}
Here, the $S/N$ corresponds to the reported detection significance, in units of $\sigma$, of the BAO feature. For further explanation of this effect and how it relates to BAO measurements, see \cite{Bassett:2010ku, Ruiz:2012rc}. Note that if $\Delta \chi^2_\text{G}$ is a combined value for multiple measurements, such as a pair of anisotropic BAO measurements, the relevant $(S/N)^2$ is the Gaussian $\Delta \chi^2$ value that corresponds to the detection \textit{probability}. For instance, ${(S/N)^2 = 6.18}$ rather than 4.00 for a 2$\sigma$ (95.4\%) detection and a $\Delta \chi^2_\text{G}$ with two degrees of freedom (see e.g.\ \cite{NumRec}).

The detection significance quoted for 6dFGS is 2.4$\sigma$. For SDSS MGS, the detection significance is roughly 2$\sigma$, but since the likelihood is non-Gaussian anyway, we apply Eq.~\eqref{eq:chi2fit} to the publically-available $\chi^2$ look-up table. We also truncate its $\Delta \chi^2$ contribution at ${\Delta \chi^2 = 3.43}$ to avoid extrapolating beyond the edge of the table. No detection significance was explicitly quoted for BOSS LOWZ, so we assume a 4$\sigma$ detection as a conservative guess. The quoted detection significance is more than 7$\sigma$ for BOSS CMASS, but we use a value of 6$\sigma$ in the analysis, in case the likelihood becomes non-Gaussian for other reasons at such a high confidence level. Finally, for BOSS Ly$\alpha$F, the detection significance is 5$\sigma$ for the auto-correlation measurement and roughly 4$\sigma$ for the cross-correlation measurement. Since these measurements are almost completely independent (see \cite{Delubac:2014aqe}), we simply add their publically-available $\chi^2$ tables. Although the combined detection significance is presumably higher, we apply Eq.~\eqref{eq:chi2fit} to the combined $\chi^2$ table assuming a significance of only 4$\sigma$. We then truncate the $\Delta \chi^2$ contribution at ${\Delta \chi^2 = 15.93}$ (less than 4$\sigma$ for two degrees of freedom) to avoid any extrapolation beyond the table. We have verified that the final results are qualitatively insensitive to the exact choices here and quantitatively sensitive to only the Ly$\alpha$F BAO significance, which we discuss in Sec.~\ref{sec:results}.

Measurements of the BAO scale are typically calibrated by the CMB, which effectively fixes the sound horizon $r_d$ by precisely constraining $\Omega_m h^2$ and $\Omega_b h^2$. Here we avoid using the CMB to measure the sound horizon, as this measurement is model-dependent. To this end, we simply allow $r_d$ to be a free parameter, effectively using only \textit{relative} distance information from BAO. This is analogous to SN Ia analysis, where one usually marginalizes over the SN Ia absolute magnitude, using only relative distance information to constrain dark energy. The CMB also provides its own precise measurement of the angular diameter distance to recombination at ${z_* \simeq 1090}$ that breaks degeneracies among the parameters describing expansion. Here we will also leave out this high-redshift distance measurement; while it can be thought of as just another BAO measurement (i.e.\ $r(z_*)/r_d$), the interpretation is not completely model-independent because the redshift $z_*$ of the CMB cannot be determined without a model. See \cite{Aubourg:2014yra} for further discussion of these different options for calibrating BAO measurements.

\section{Methodology} \label{sec:method}

In this section, we review the statistics we use to determine goodness-of-fit and perform the model comparison.

\subsection{Goodness of fit}

To determine if a model is a good fit to the data, we minimize $\chi^2$ over the free model parameters and calculate the probability $P(\chi^2_\text{min}, \nu)$ that a greater $\chi^2_\text{min}$ could occur due to chance alone for a fit with $\nu = N - k$ degrees of freedom, where $N$ is the total number of measurements and $k$ is the number of free model parameters. This probability is given by
\begin{equation}
P(\chi^2, \nu) = \frac{\displaystyle \Gamma \left(\frac{\nu}{2}, \frac{\chi^2}{2} \right)}{\displaystyle \Gamma \left(\frac{\nu}{2} \right)} ,
\end{equation}
where $\Gamma(s, x)$ is the upper incomplete gamma function,
\begin{equation}
\Gamma(s, x) = \int_x^\infty t^{s - 1} e^{-t} dt \ ,
\end{equation}
and $\Gamma(s) = \Gamma(s, 0)$ is the (complete) gamma function.

\subsection{Model comparison}

We will use three alternative methods for model comparison. Formally, the different statistics have different meanings and are valid under different assumptions, so obtaining consistent results using different criteria mitigates the possibility that invalid assumptions about the nature of the data will favor one model over another and lead to invalid conclusions. Each of these statistics accounts for the fact that a simpler model (one with fewer free parameters) is preferable to a more complex model if both fit the data similarly well. Note that they do not require the different models to be nested; while $\rhct$ is nested within power law cosmology, neither is nested with $\Lambda$CDM. For more information about these statistics, and for other interesting uses of model comparison in cosmology, see \cite{Liddle:2004nh, Trotta:2005ar, Mukherjee:2005tr, Trotta:2008qt, Verde:2013cqa, Verde:2013wza, Leistedt:2014sia}.

The Akaike information criterion (AIC) \cite{Akaike:AIC}, which is grounded in information theory, estimates how much more information is lost when describing data with one model over another. For a best-fit $\chi^2_\text{min}$ and a model with $k$ free parameters, AIC is given by
\begin{equation}
\text{AIC} = \chi^2_\text{min} + 2 k \ .
\end{equation}
This is an asymptotic expression, and a second-order correction term can be added to make the criterion more accurate for a finite number of observations:
\begin{equation}
\text{AICc} = \text{AIC} + \frac{2 k \left(k + 1 \right)}{N - k - 1} \ .
\end{equation}
This makes a small difference when the number of data points $N$ is large (e.g.\ for the SN data) but a significant difference if $N$ is small (e.g.\ for the BAO data). Since AICc reduces to AIC in the limit of large $N$, we use AICc instead of AIC throughout the analysis.

The Bayesian information criterion (BIC) \cite{Schwarz:BIC} is also an asymptotic expression, and it follows from a Bayesian argument that considers likelihoods in the exponential family of probability distributions, which includes the Gaussian distribution and many other common distributions. BIC selects the model that is a posteriori most probable. It is given by
\begin{equation}
\text{BIC} = \chi^2_\text{min} + k \ln(N) \ .
\end{equation}
BIC typically (though not always) penalizes extra parameters more severely than AIC.

The Bayes factor $B_{1 0}$ indicates the likelihood of one model relative to another by integrating both likelihoods over all values of the model parameters, weighting them by the priors. This statistic is presumably the most robust, as it considers all values, not just the best-fit values, of the parameters. It naturally penalizes a model with more free parameters, especially if those parameters do not lead to a better fit. In fact, BIC can be considered an approximation to the logarithm of the Bayes factor. For a set of data $D$ and two different models $M_0$ and $M_1$ that are described, respectively, by sets of parameters $\bm{\theta}_0$ and $\bm{\theta}_1$, the Bayes factor indicates the likelihood of $M_1$ relative to $M_0$ and is given by
\begin{equation}
B_{1 0} = \frac{\displaystyle \int \text{Pr}(D | \bm{\theta}_1, M_1) \ \text{Pr}(\bm{\theta}_1 | M_1) \ d\bm{\theta}_1}{\displaystyle \int \text{Pr}(D | \bm{\theta}_0, M_0) \ \text{Pr}(\bm{\theta}_0 | M_0) \ d\bm{\theta}_0} \ .
\end{equation}
For this analysis, we take the prior distributions $\text{Pr}(\bm{\theta} | M)$ to be flat, and we assume that the likelihoods $\text{Pr}(D | \bm{\theta}, M)$ are Gaussian (this assumption is implicit in our definitions of AIC and BIC, where we write $\chi^2_\text{min}$ in place of the more general $-2 \ln(\mathcal{L}_\text{max})$). We compute the likelihoods numerically over grids of parameter values. Analytic marginalization over $\mathcal{M}$ and $\Delta M$ (e.g.\ Appendix of \cite{Shafer:2013pxa}) leaves at most four parameters over which to grid, making this brute-force approach feasible.

\section{Results} \label{sec:results}

Table~\ref{tab:bestfit} lists the best-fit values for the model parameters, including the nuisance parameters for the SN and BAO data, for each model discussed in Sec.~\ref{sec:models} and data combination discussed in Sec.~\ref{sec:data}. Table~\ref{tab:compare} shows the results of the model comparison. For each model and data combination, we list the number of parameters $k$ that were varied, the total number of data points $N$, the best-fit $\chi^2_\text{min}$, the probability $P(\chi^2_\text{min}, \nu)$ that a greater $\chi^2_\text{min}$ could occur due to chance alone for degrees of freedom ${\nu = N - k}$, and the likelihood of the model relative to $\Lambda$CDM for the AICc, BIC, and Bayes factor model comparison statistics. We note that, while statisticians have proposed various scales that give a qualitative interpretation of the numerical results of model comparison statistics (e.g.\ Jeffreys' scale for the Bayes factor), these are obviously rather subjective, so here we simply discuss the results in terms of relative probabilities (e.g.\ $\exp(-\Delta \text{BIC} /2)$) and let the reader judge their significance.

From Table~\ref{tab:compare}, it is clear that $\Lambda$CDM is a good fit to all of the SN Ia data. When systematic errors are included, the fit may even be slightly \textit{too} good, indicating that the errors may be overestimated. This would not be surprising, as the SN analyses generally aim to be conservative when estimating the magnitude of systematic errors. A more surprising result is that $\Lambda$CDM is actually not a very good fit to the BAO data. The fit, with a probability of 0.052, corresponds to a nearly 2$\sigma$ discrepancy. This tension has already been noted (e.g.\ \cite{Aubourg:2014yra}) and is due to the anisotropic Ly$\alpha$F BAO measurements, with the radial measurement $c/(H(2.34) \, r_d)$ too high and the transverse measurement $r(2.34)/r_d$ too low for $\Lambda$CDM. This is the case regardless of whether $\Omega_m$ and $r_d$ are constrained by CMB observations (e.g.\ \textit{Planck}) or by the combined set of BAO measurements, as in this analysis. The similar value of $\Omega_m$ preferred by both SN Ia and BAO data means extra tension will not arise when combining the data, and so the combined SN~+~BAO fits are still very good.

\begin{table*}
\setlength{\tabcolsep}{0.6em}
\begin{tabular}{llccccccr}
\hline \hline
Model & Data & $\Omega_m$ & $n$ & $r_d \times H_0/c$ & $\alpha$ & $\beta$ & $\mathcal{M}$ & $\Delta M$ \\ \hline
$\Lambda$CDM & JLA (Stat) & 0.287 & - & - & 0.140 & 3.14 & 24.11 & -0.060 \\
$\Lambda$CDM & JLA (Sys) & 0.294 & - & - & 0.141 & 3.10 & 24.11 & -0.070 \\
$\Lambda$CDM & Union2.1 (Stat) & 0.278 & - & - & - & - & 43.16 & 0.000 \\
$\Lambda$CDM & Union2.1 (Sys) & 0.295 & - & - & - & - & 43.17 & 0.000 \\
$\Lambda$CDM & BAO & 0.285 & - & 0.0338 & - & - & - & - \\
$\Lambda$CDM & BAO + JLA (Stat) & 0.286 & - & 0.0338 & 0.140 & 3.14 & 24.11 & -0.059 \\
$\Lambda$CDM & BAO + JLA (Sys) & 0.288 & - & 0.0338 & 0.141 & 3.10 & 24.11 & -0.070 \\
$\Lambda$CDM & BAO + Union2.1 (Stat) & 0.282 & - & 0.0339 & - & - & 43.16 & -0.002 \\
$\Lambda$CDM & BAO + Union2.1 (Sys) & 0.288 & - & 0.0338 & - & - & 43.16 & 0.001 \\ \hline
Power Law & JLA (Stat) & - & 1.56 & - & 0.139 & 3.14 & 24.14 & -0.061 \\
Power Law & JLA (Sys) & - & 1.55 & - & 0.141 & 3.10 & 24.13 & -0.071 \\
Power Law & Union2.1 (Stat) & - & 1.54 & - & - & - & 43.20 & -0.022 \\
Power Law & Union2.1 (Sys) & - & 1.44 & - & - & - & 43.20 & -0.003 \\
Power Law & BAO & - & 0.93 & 0.0301 & - & - & - & - \\
Power Law & BAO + JLA (Stat) & - & 1.52 & 0.0333 & 0.139 & 3.13 & 24.14 & -0.062 \\
Power Law & BAO + JLA (Sys) & - & 1.45 & 0.0331 & 0.140 & 3.09 & 24.14 & -0.072 \\
Power Law & BAO + Union2.1 (Stat) & - & 1.49 & 0.0332 & - & - & 43.20 & -0.027 \\
Power Law & BAO + Union2.1 (Sys) & - & 1.35 & 0.0326 & - & - & 43.21 & -0.007 \\ \hline
$R_\text{h} = ct$ & JLA (Stat) & - & - & - & 0.131 & 3.13 & 24.24 & -0.083 \\
$R_\text{h} = ct$ & JLA (Sys) & - & - & - & 0.138 & 3.07 & 24.23 & -0.077 \\
$R_\text{h} = ct$ & Union2.1 (Stat) & - & - & - & - & - & 43.33 & -0.115 \\
$R_\text{h} = ct$ & Union2.1 (Sys) & - & - & - & - & - & 43.29 & -0.026 \\
$R_\text{h} = ct$ & BAO & - & - & 0.0308 & - & - & - & - \\
$R_\text{h} = ct$ & BAO + JLA (Stat) & - & - & 0.0308 & 0.131 & 3.13 & 24.24 & -0.083 \\
$R_\text{h} = ct$ & BAO + JLA (Sys) & - & - & 0.0308 & 0.138 & 3.07 & 24.23 & -0.077 \\
$R_\text{h} = ct$ & BAO + Union2.1 (Stat) & - & - & 0.0308 & - & - & 43.33 & -0.115 \\
$R_\text{h} = ct$ & BAO + Union2.1 (Sys) & - & - & 0.0308 & - & - & 43.29 & -0.026 \\
\hline \hline
\end{tabular}
\caption{Best-fit values of each parameter varied for each model and data combination.}
\label{tab:bestfit}
\end{table*}

\begin{table*}
\setlength{\tabcolsep}{0.6em}
\begin{tabular}{llllrllll}
\hline \hline
Model & Data & $k$ & $N$ & $\chi^2_\text{min}$ & $P(\chi^2_\text{min}, \nu)$ & $\exp(-\Delta \text{AICc} /2)$ & $\exp(-\Delta \text{BIC} /2)$ & $B_{1 0}$ \\ \hline
$\Lambda$CDM & JLA (Stat) & 5 & 740 & 722.6 & 0.62 & 1 & 1 & 1 \\
$\Lambda$CDM & JLA (Sys) & 5 & 740 & 682.7 & 0.92 & 1 & 1 & 1 \\
$\Lambda$CDM & Union2.1 (Stat) & 3 & 580 & 562.2 & 0.66 & 1 & 1 & 1 \\
$\Lambda$CDM & Union2.1 (Sys) & 3 & 580 & 545.1 & 0.83 & 1 & 1 & 1 \\
$\Lambda$CDM & BAO & 2 & 7 & 11.0 & 0.052 & 1 & 1 & 1 \\
$\Lambda$CDM & BAO + JLA (Stat) & 6 & 747 & 733.6 & 0.57 & 1 & 1 & 1 \\
$\Lambda$CDM & BAO + JLA (Sys) & 6 & 747 & 693.8 & 0.89 & 1 & 1 & 1 \\
$\Lambda$CDM & BAO + Union2.1 (Stat) & 4 & 587 & 573.3 & 0.61 & 1 & 1 & 1 \\
$\Lambda$CDM & BAO + Union2.1 (Sys) & 4 & 587 & 556.1 & 0.78 & 1 & 1 & 1 \\ \hline
Power Law & JLA (Stat) & 5 & 740 & 725.8 & 0.59 & 0.20 & 0.20 & 0.17 \\
Power Law & JLA (Sys) & 5 & 740 & 682.7 & 0.92 & 1.0 & 1.0 & 0.84 \\
Power Law & Union2.1 (Stat) & 3 & 580 & 567.6 & 0.60 & 0.069 & 0.069 & 0.057 \\
Power Law & Union2.1 (Sys) & 3 & 580 & 547.0 & 0.81 & 0.39 & 0.39 & 0.28 \\
Power Law & BAO & 2 & 7 & 8.8 & 0.12 & 3.0 & 3.0 & 1.3 \\
Power Law & BAO + JLA (Stat) & 6 & 747 & 748.7 & 0.41 & 0.00053 & 0.00053 & 0.00065 \\
Power Law & BAO + JLA (Sys) & 6 & 747 & 705.0 & 0.82 & 0.0037 & 0.0037 & 0.0052 \\
Power Law & BAO + Union2.1 (Stat) & 4 & 587 & 590.1 & 0.41 & 0.00022 & 0.00022 & 0.00021 \\
Power Law & BAO + Union2.1 (Sys) & 4 & 587 & 567.9 & 0.67 & 0.0028 & 0.0028 & 0.0026 \\ \hline
$R_\text{h} = ct$ & JLA (Stat) & 4 & 740 & 855.8 & 0.0014 & 3.2$\times 10^{-29}$ & 3.2$\times 10^{-28}$ & 1.0$\times 10^{-28}$ \\
$R_\text{h} = ct$ & JLA (Sys) & 4 & 740 & 721.3 & 0.64 & 1.2$\times 10^{-8}$ & 1.1$\times 10^{-7}$ & 2.0$\times 10^{-8}$ \\
$R_\text{h} = ct$ & Union2.1 (Stat) & 2 & 580 & 656.7 & 0.013 & 8.3$\times 10^{-21}$ & 7.2$\times 10^{-20}$ & 2.3$\times 10^{-20}$ \\
$R_\text{h} = ct$ & Union2.1 (Sys) & 2 & 580 & 565.7 & 0.63 & 9.2$\times 10^{-5}$ & 0.00081 & 0.00013 \\
$R_\text{h} = ct$ & BAO & 1 & 7 & 18.3 & 0.0056 & 0.21 & 0.069 & 0.42 \\
$R_\text{h} = ct$ & BAO + JLA (Stat) & 5 & 747 & 874.1 & 0.00055 & 8.3$\times 10^{-31}$ & 8.3$\times 10^{-30}$ & 1.9$\times 10^{-30}$ \\
$R_\text{h} = ct$ & BAO + JLA (Sys) & 5 & 747 & 739.6 & 0.52 & 3.1$\times 10^{-10}$ & 3.0$\times 10^{-9}$ & 5.1$\times 10^{-10}$ \\
$R_\text{h} = ct$ & BAO + Union2.1 (Stat) & 3 & 587 & 675.0 & 0.0053 & 2.2$\times 10^{-22}$ & 2.0$\times 10^{-21}$ & 4.9$\times 10^{-22}$ \\
$R_\text{h} = ct$ & BAO + Union2.1 (Sys) & 3 & 587 & 584.0 & 0.49 & 2.5$\times 10^{-6}$ & 2.2$\times 10^{-5}$ & 4.1$\times 10^{-6}$ \\
\hline \hline
\end{tabular}
\caption{Results of the model comparison. For each of the models and data combinations, we list the number of parameters $k$ that were varied, the total number of data points $N$, the best-fit $\chi^2_\text{min}$, the probability $P(\chi^2_\text{min}, \nu)$ that a greater $\chi^2_\text{min}$ could occur due to chance alone for degrees of freedom ${\nu = N - k}$, and the likelihood of the model relative to $\Lambda$CDM for the AICc, BIC, and Bayes factor model comparison statistics.}
\label{tab:compare}
\end{table*}

As we will see below, the uncalibrated galaxy BAO distances (our set without the Ly$\alpha$F BAO) are not effective in distinguishing between the expansion models we consider here. Although the galaxy BAO measurements are more mature and their systematics have been more thoroughly studied, there is no obvious reason why we should ignore the Ly$\alpha$F measurements. The original analyses \cite{Delubac:2014aqe, Font-Ribera:2014wya} investigate some important systematic effects, and they find no substantial evidence that contamination from these systematics is large. They also show that the measurements are generally robust to variations in the fiducial analysis pipeline. We therefore use the full set of BAO measurements, including the high-redshift Ly$\alpha$F BAO, leaving open the possibility that power law or $\rhct$ cosmology improves the fit to the anisotropic data.

\begin{figure}
\includegraphics[width=\linewidth]{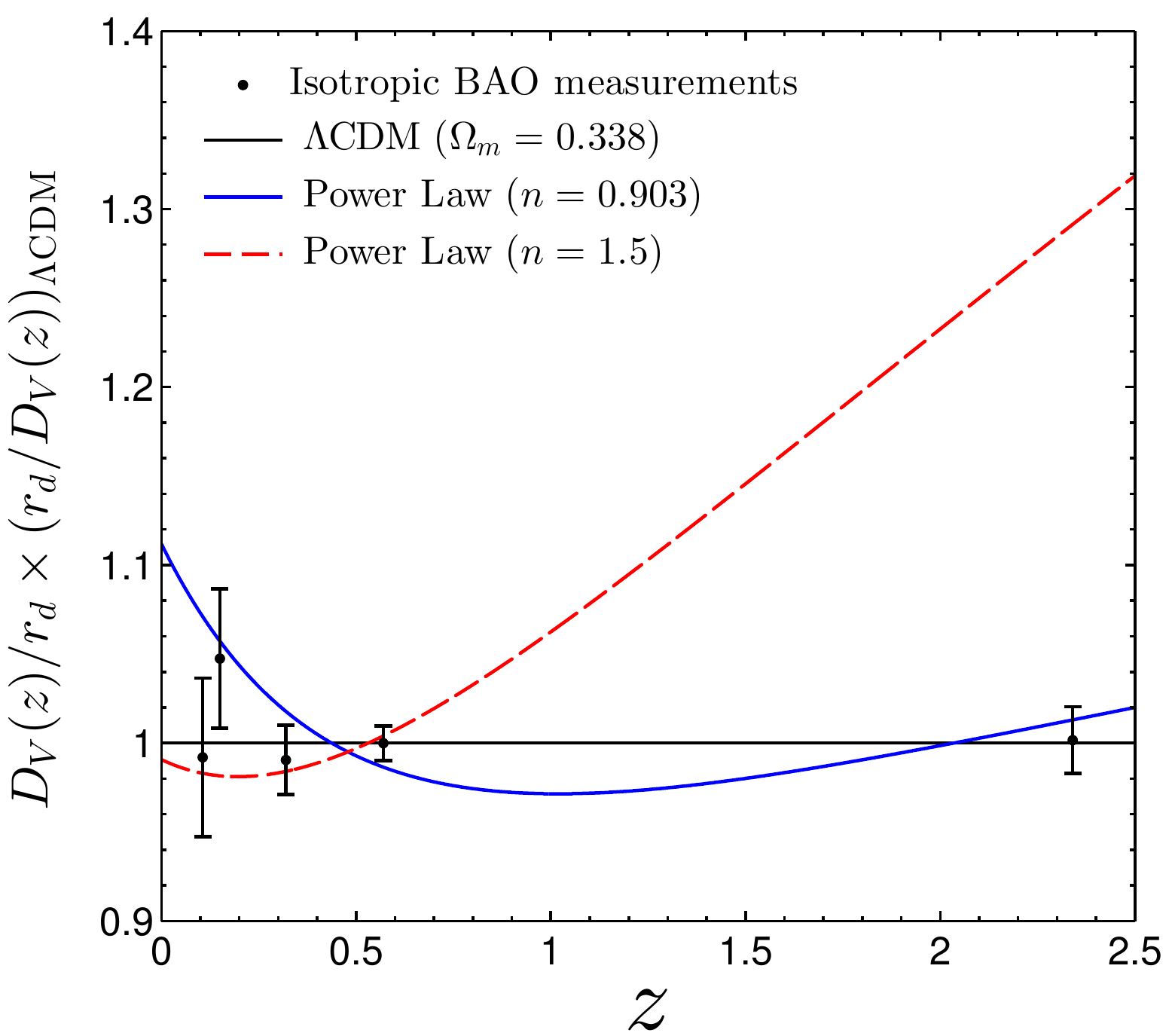}
\caption{Fits to an isotropic-only version of the BAO data (black points), where we use the direct isotropic measurement from BOSS CMASS (${z = 0.57}$) and an isotropic measurement derived from the Ly$\alpha$F anisotropic measurements (${z = 2.34}$). We show the best fit to this modified BAO set for $\Lambda$CDM with $\Omega_m$ and $r_d$ varied (solid black), power law cosmology with $n$ and $r_d$ varied (solid blue), and power law cosmology with ${n = 1.5}$ and only $r_d$ varied (dashed red), where the value ${n = 1.5}$ is roughly the value required to fit the SN Ia data.}
\label{fig:bao}
\end{figure}

Interestingly, power law cosmology is about as good a fit to the SN and BAO data separately as $\Lambda$CDM. If systematic errors in the SN data are ignored, $\Lambda$CDM is preferred, but when the systematics are included, the preference for $\Lambda$CDM nearly disappears. Here the AICc and BIC statistics give identical results simply because $\Lambda$CDM and power law cosmology both have one free model parameter. Power law expansion is actually a better fit to the BAO data, though the model comparison statistics indicate only a mild preference for power law expansion. The Bayes factor, which considers the likelihood averaged over the respective parameter spaces, is particularly indifferent.

The story changes when SN and BAO data are combined, however, and here $\Lambda$CDM is strongly favored by the data. As Table~\ref{tab:bestfit} suggests, the power law exponent preferred by SN data (${n \simeq 1.5}$) is much higher than that preferred by BAO data (${n = 0.93}$), and this tension means that the combination strongly disfavors power law cosmology relative to $\Lambda$CDM, even though power law expansion is still a good fit overall. The relative probability that power law cosmology is the ``correct'' model is at most ${0.0052 \simeq 1/200}$, which occurs for the Bayes factor when the JLA SN compilation is used with systematic errors included in the analysis.

One might wonder whether the low power law exponent (${n = 0.93}$) preferred by the BAO data is due to some effect specific to the anisotropic measurements and the extra degree of freedom they probe. To investigate this, and to visualize the fit to the BAO data, we plot an isotropic-only version of the BAO data in Fig.~\ref{fig:bao}. Here we have used the direct isotropic measurement from BOSS DR11 CMASS \cite{Anderson:2013zyy}, ${D_V(0.57)/r_d = 13.773 \pm 0.134}$, and an isotropic measurement we derived from the Ly$\alpha$F anisotropic measurements, ${D_V(2.34)/r_d = 30.543 \pm 0.570}$, where we have propagated the errors and accounted for their correlation. Note that, while the direct isotropic measurement from CMASS is in excellent agreement with a measurement inferred in this way, the result is approximately valid only if systematic errors in the anisotropic measurements are negligible compared to the statistical errors. Obviously there is reason to worry that this is \textit{not} the case for the Ly$\alpha$F BAO, so our fiducial analysis uses the anisotropic measurements directly, as recommended in the original analyses.

This issue aside, it is clear from Fig.~\ref{fig:bao} that the preference for a low power law exponent is not some artifact of the anisotropic measurements. Here we show fits to the isotropic-only BAO data for both $\Lambda$CDM and power law cosmology. Power law cosmology is a slightly worse fit than in the fiducial analysis, with ${\chi^2_\text{min} = 7.03}$ (a probability of 0.071 for three degrees of freedom). $\Lambda$CDM is a good fit now that the tension from the anisotropic Ly$\alpha$F measurements has effectively canceled, with ${\chi^2_\text{min} = 1.65}$ (a probability of 0.65 for three degrees of freedom). We also plot a power law model with the exponent fixed to the value ${n = 1.5}$, roughly the value required by the SN Ia data, adjusting only the sound horizon $r_d$ to give the best fit. Since we have assumed the same detection significances (6$\sigma$ for CMASS and 4$\sigma$ for Ly$\alpha$F) as in the fiducial analysis, it is no surprise that a smaller value for the sound horizon, which raises the model relative to the data, is preferred. Such a model fits the low-redshift BAO data nicely but misses the Ly$\alpha$F measurement completely. Increasing the detection significance of the Ly$\alpha$F measurement would make this discrepancy with the SN Ia data even more significant, and we believe that our choice here of 4$\sigma$ (effectively less when using the $\chi^2$ table for the anisotropic measurements) for the combined auto-correlation and cross-correlation measurement is conservative.

Focusing separately on the $\rhct$ universe, we find that it is conclusively disfavored relative to $\Lambda$CDM for all data combinations that include SN Ia data, with at most a relative probability of ${0.00081 \simeq 1/1200}$, which occurs for BIC when the Union2.1 SN compilation is used without BAO data and with systematic errors included. If systematic errors in the SN data are ignored, the $\rhct$ universe is a poor fit to the data (for JLA, the fit probability of 0.0014 corresponds to a $>3\sigma$ discrepancy). The $\rhct$ universe is also a poor fit to the BAO data alone and, despite the fact that $\rhct$ benefits from having one less free parameter, $\Lambda$CDM is slightly preferred by the model comparison statistics. Note that here, with $N$ very small, we do not expect the BIC result to be valid, and indeed we find that the AICc statistic, with its correction term for finite sample size, is closer to the Bayes factor result.

\begin{figure*}
\includegraphics[width=\linewidth]{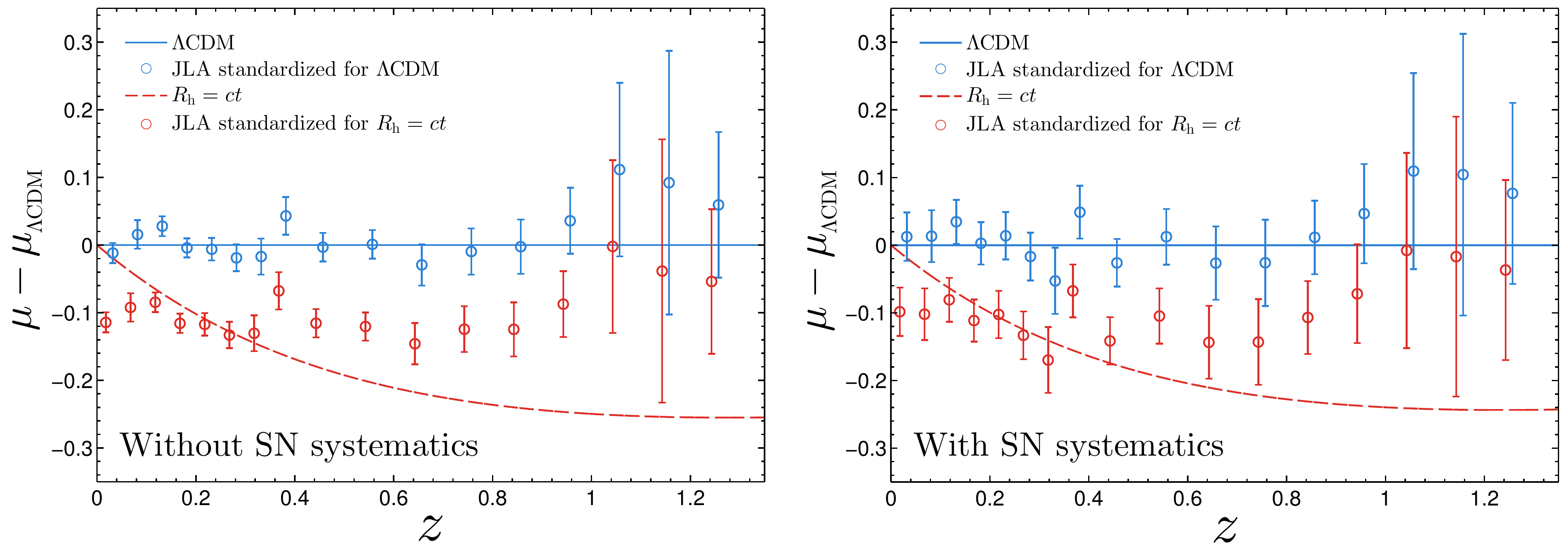}
\caption{Hubble diagram for the JLA SN compilation, where the measured distance moduli have been standardized separately for $\Lambda$CDM (blue points) and the $\rhct$ cosmology (red points). The best-fit (${\Omega_m = 0.29}$) $\Lambda$CDM model (solid blue) is plotted along with the $\rhct$ model (dashed red). We show the SN data without (left panel) and with (right panel) systematic errors included. The distance moduli are binned in redshift with inverse-covariance weights.}
\label{fig:hubble}
\end{figure*}

Figure~\ref{fig:hubble} illustrates the $\Lambda$CDM and $\rhct$ fits to the JLA SN data, which we have binned in redshift by averaging the distance moduli with inverse-covariance weights. There are two sets of data points because the SN data were standardized separately for each model, with $\mathcal{M}$, $\Delta M$, $\alpha$, and $\beta$ optimized to produce the best fit. While $\Lambda$CDM is clearly a good fit with or without systematic errors included, it is apparent by eye that $\rhct$ is a poor fit with statistical errors only. While the overall fit looks reasonable when systematic errors are included, there is a clear trend in the residuals, with nearby SNe too bright and distant SNe too dim.

\section{Discussion} \label{sec:discuss}

In this analysis, we have used goodness-of-fit and model comparison statistics to test power law and $\rhct$ cosmology against $\Lambda$CDM. Before we summarize our conclusions, we discuss some arguments made by proponents of the $\rhct$ universe \cite{Melia:2012zy, Melia:2013hsa, Wei:2015xca}, particularly those about hidden model dependence in the SN Ia data.

As mentioned in Sec.~\ref{sec:intro}, the authors of \cite{Melia:2013hsa} criticize the analysis of \citet{Bilicki:2012ub}, pointing out that the conclusions are based heavily on visual inspection of plots of reconstructed dynamical quantities (e.g.\ derivatives of the Hubble parameter). While there is nothing wrong with a dynamical approach, one would like a way to quantify the preference of the data. Here we have used direct model comparison to quantify the relative likelihoods of the alternative models, appropriately accounting for differences in model complexity. Note that a similar approach is used in \cite{Wei:2015xca}.

One of the concerns \cite{Melia:2012zy} about the SN Ia data is that the standardization of the SNe is model-dependent, in the sense that the parameters $\alpha$, $\beta$, $\mathcal{M}$, and $\Delta M$ have been determined assuming $\Lambda$CDM (or a $\Lambda$CDM-like model). This is a valid point in general, but only if one analyzes distance moduli that have been pre-corrected for stretch, color, and host-mass correlations. Fixing the SN Ia absolute magnitude and Hubble constant (thus fixing $\mathcal{M}$) would be even worse. In this analysis, we have used the public JLA data to vary all of the nuisance parameters simultaneously with any cosmological parameters (\cite{Wei:2015xca} does something similar using an older version of the SNLS data). It is worth noting that, aside from adjusting the overall offset in the Hubble diagram $\mathcal{M}$, allowing the other parameters to vary has a rather small effect. This is hinted at in Table~\ref{tab:bestfit}, where the best-fit values for $\alpha$, $\beta$, and even $\Delta M$ are similar for the different models and regardless of whether systematic errors are included. This is because these correlations, which have now been well-established, primarily serve to reduce the intrinsic scatter in the Hubble diagram. Only if there is, for instance, a trend where the \textit{fraction} of SNe with high-stretch light curves changes significantly with redshift would the choice of model affect the value of $\alpha$ significantly. Note that the SN analyses do consider possible evolution of $\alpha$ and $\beta$ and account for this by adding systematic error.

The authors of \cite{Wei:2015xca} go further and argue that, because the intrinsic scatter is determined by adjusting its value until the reduced $\chi^2$ of the fit is equal to one, the intrinsic scatter estimate is model-dependent. This concern is actually addressed in Section~5.5 of the JLA analysis \cite{Betoule:2014frx}, which explains that one can avoid this problem by estimating the intrinsic scatter in redshift bins, essentially relying on the fact that the SNe are so constraining, and the redshift coverage so complete, that no parametric model needs to be used at all. In other words, the scatter around the mean in a redshift bin gives the intrinsic scatter for that bin directly. Although \cite{Betoule:2014frx} finds an apparent trend where the estimated intrinsic scatter decreases with increasing redshift, the values are consistent with a constant, and ultimately a separate value is chosen for each of the four main subsamples, effectively allowing for survey-dependent misestimates of other statistical errors. We point out that the approach used in \cite{Wei:2015xca}, where $\sigma_\text{int}$ is constrained along with the other parameters, results in nearly identical determinations of the intrinsic scatter whether assuming $\Lambda$CDM (${\sigma_\text{int} = 0.103 \pm 0.010}$) or $\rhct$ (${\sigma_\text{int} = 0.106 \pm 0.010}$), indicating that even if the model matters in general, it does not in their case.

Why does \cite{Wei:2015xca} come to a different conclusion (that $\rhct$ is modestly favored over $\Lambda$CDM) than the present analysis, where similar SN Ia data is used? Presumably the reason is that \cite{Wei:2015xca} uses \textit{only} the SNLS SNe, leaving out the low-redshift samples, the mid-redshift SDSS SNe, and the high-redshift SNe from the Hubble Space Telescope, without any real justification. This of course removes much of the discriminating power of the SN data. In Fig.~\ref{fig:hubble}, notice that only considering the data points in the range ${0.4 < z < 1}$, which roughly corresponds to the redshift range dominated by the SNLS SNe, would make $\rhct$ a great fit after the overall height of the data points ($\mathcal{M}$) is adjusted. Even the trend in these residuals (presumably due to their statistical correlation) aligns with the $\rhct$ expansion. While it is true that combining observations of SNe from different instruments into one Hubble diagram is challenging and can lead to concerns about residual systematic effects, the most important systematic errors (e.g.\ photometric calibration) are quantified in current analyses. For some reason, despite their apparent worry about unaccounted-for systematic effects, the authors of \cite{Wei:2015xca} ignore the \textit{known} systematic errors altogether, a common trend in the literature that is usually unjustified and will lead to false conclusions. Here, for instance, systematic errors substantially weaken the model-discriminating power of the SN data; in fact, it is these systematic errors that allow the $\rhct$ universe to be an acceptable overall fit, even if it is conclusively disfavored by the model comparison.

These concerns notwithstanding, we are confident in our conclusions, which we summarize as follows. Given the tension in the BAO data resulting from the anisotropic Ly$\alpha$F measurements, we find that power law cosmology is a slightly better model than $\Lambda$CDM for BAO data alone and is only slightly disfavored relative to $\Lambda$CDM for SN Ia data alone. When SN and BAO data are combined, the different power law exponents preferred by each create substantial tension such that $\Lambda$CDM is strongly preferred by the model comparison statistics (Table~\ref{tab:compare}). While the strength of this preference depends on the detection significance of the BOSS Ly$\alpha$F measurement, our choice here (see Sec.~\ref{sec:BAO}) is conservative. Also, while the preference for a low power law exponent (${n \simeq 0.9}$) is due to the Ly$\alpha$F BAO, it is not due to any subtlety in the anisotropic measurements (see Fig.~\ref{fig:bao} and the discussion in Sec.~\ref{sec:results}). In order to reconcile the power law exponent from BAO with the higher value (${n \simeq 1.5}$) required by SN Ia data, an unaccounted-for systematic affecting the anisotropic Ly$\alpha$F measurements must avoid cancellation in the isotropic measurement, shifting the value by several times its current error. Hopefully, new data (e.g.\ BOSS DR12) will soon improve these high-redshift BAO measurements, which are clearly important for distinguishing between alternative models for expansion, including those considered here.

We have also found that the $\rhct$ cosmology is nearly ruled out when systematic errors in the SN Ia data are ignored (for JLA, the fit is unlikely at a level of $>3\sigma$). With systematic errors included, the overall fit is acceptable, but $\Lambda$CDM is conclusively the better model. The BAO data separately favor $\Lambda$CDM, though only slightly, and combining SN and BAO data simply strengthens these conclusions.

\vfill

\begin{acknowledgments}

I thank Dragan Huterer and the anonymous referee for valuable discussions and thoughtful comments on the manuscript. This work has been supported by DOE Grant No.\ DE-FG02-95ER40899 and NSF Grant No.\ AST-0807564.

\end{acknowledgments}

\bibliography{powerlaw}

\end{document}